\documentclass{wqcd03}                 

\newcommand{\nn}{\nonumber}
\newcommand{\IM}{\mbox{\rm Im}}

\newcommand{\mev}{\mbox{\rm MeV}}
\newcommand{\gev}{\mbox{\rm GeV}}

\newcommand{\eqn}[1]{(\ref{#1})}

\newcommand{\MSbtext}{$\overline{\rm MS}$}
\newcommand{\ep}{\epsilon}

\usepackage{txfonts}                   
\confname{QCD@Work 2003 - International Workshop on QCD, Conversano, Italy,
14--18 June 2003}

\title{{\small \hfill IFIC/03-46, HD-THEP-03-52, MPP-2003-110}\\
Finite energy sum rules for the vector current: the strange quark mass}

\author{Markus~Eidem\"uller\addressmark{a}\thanks{Speaker at the Workshop.}, 
Matthias~Jamin\addressmark{b} and Felix~Schwab\addressmark{c}}

\address[a]{Departament de F\'{\i}sica Te\`orica, IFIC, Universitat de
  Val\`encia -- CSIC, Apt. Correus 22085, E-46071 Val\`encia, Spain}
\address[b]{Institut f\"ur Theoretische Physik, Universit\"at Heidelberg, 
Philosophenweg 16, D-69120 Heidelberg, Germany}
\address[c]{Max-Planck-Institut f\"ur Physik -- Werner-Heisenberg-Institut,
F\"ohringer Ring 6, D-80805 M\"unchen, Germany}

\begin{document}

\begin{abstract}
We determine the strange quark mass in the framework of finite energy sum rules
from the vector current channel. The theoretical contributions are calculated
in contour improved perturbation theory and a substantial difference to 
fixed order perturbation theory is found.
In our phenomenological parametrisation we include recent experimental results 
from CMD-2. 
\end{abstract}

\maketitle



\section{Introduction}

During the last years, the strange quark mass has been the subject of
intense investigation. This development was driven by the aim to improve the
theoretical value of $\ep'/\ep$, 
where the strange quark mass represents one of the dominant uncertainties.
More recently, it was found that in predictions
of the QCD factorization approach to matrix elements for weak decays
\cite{Beneke:2003zv}
and in estimates of SU(3) breaking effects \cite{Khodjamirian:2003xk}
large uncertainties arise due to the value of the strange quark mass.

In order to extract the
strange quark mass, QCD sum rules \cite{SVZ,RRY:1985,Narison:1989} have  
been applied to different
channels. Recent progress has been made in the sum rules for hadronic
$\tau$-decays \cite{Gamiz:2002nu,mutau}. In the scalar channel the work of 
\cite{Colangelo:1997uy} has lately been updated in \cite{JOP2} with a substantial
improvement in the phenomenological description. The pseudoscalar sum rules 
have been investigated in \cite{MK}. In \cite{Korner:2000wd} the strange quark 
mass was determined from Cabibbo-suppressed $\tau$-decays. 
In this work, we intend to clarify the 
situation in the vector channel which allows to obtain the 
strange quark mass from the cross section for $e^{+}e^{-}$-scattering.

A $\tau$-like finite energy sum rule in the vector channel was first introduced 
by Narison \cite{Narison95}. The analysis used fixed order perturbation theory (FOPT)
for the isovector minus isoscalar current and an \MSbtext-mass of 
$m_s(2\ \gev)=143\pm 21\ \mev$ was obtained. In \cite{Maltman98,Maltman:1998xn}
Maltman pointed out
a possible large contribution from $\rho-\omega$ mixing to this sum rule which
reduced the mass to $m_s(2\ \gev)=97\pm 50\ \mev$ with a huge error. Later,
Narison presented an updated analysis \cite{Narison99} where he introduced a
new sum rule for the strange quark current which is free from
$\rho-\omega$ mixing and found  $m_s(2\ \gev)=129\pm 24\ \mev$. In view of this
situation, we considered it important to perform an independent analysis in this
channel \cite{EJS}. 
In our theoretical expressions, contour improved perturbation theory
(CIPT)\cite{Pivovarov:1991rh,DP,PP1} has been used. This method improves the convergence 
of the perturbative series and was successfully applied to $\tau$-decays.
Surprisingly, we find a large difference between the CIPT and FOPT evaluations 
in the leading
mass corrections and we comment on this topic in section 2.3. This leads to a
substantial shift in the strange quark mass with respect to the FOPT determinations.
For the phenomenological parametrisation in our sum rules we use
the new results from the Novosibirsk-CMD-2 experiment for the $\phi$-resonance
\cite{CMD2}.


\section{Finite energy sum rules}


\subsection{Definitions}

The central object in this investigation is the vector current two-point correlator 
\begin{eqnarray}
  \label{eq:2.a}
        \Pi_{\mu\nu}(q^2) &=& i \int d^4 x \ e^{iqx}\, \langle
        T\{j_\mu(x) j_\nu^\dagger(0)\}\rangle \nn\\ 
          &=& (q_\mu q_\nu-g_{\mu\nu}q^2)\,\Pi(q^2)\,.
\end{eqnarray}
In our case we are interested in the sum rule for the vector strange current
which is given by $j_\mu(x)= e_s \bar{s}(x)\gamma_\mu s(x)$ and
$e_s=-1/3$ is the electric charge of the strange quark.
Via the optical theorem, the imaginary part of the correlator is related to the
cross section of strange quark production
\begin{center}
\begin{equation}
  \label{eq:2.b}
          R(q^2)=\frac{\sigma(e^+ e^- \to \gamma^*(q) \to s\bar{s})}
        {\sigma(e^+ e^- \to \mu^+ \mu^-)}=12\pi\,\IM\, \Pi(q^2+i\ep)\,.
\end{equation}
\end{center}
In the finite energy sum rules one defines the moments by an integral of
$\Pi(s)$ along a circle with radius $r^2$,
\begin{eqnarray}
 \label{eq:2.c}
{\mathcal M}^{k}(r) &=& 6 \pi i\oint_{|s|=r^2} \frac{ds}{r^2}\ \Pi(s)w_r^{k}(s)\nn\\
&=& \int_{0}^{r^2} \frac{ds}{r^2}\ R(s)w_r^{k}(s)\,,\quad\mbox{with}\nn\\
w_r^{k}(s)&=& \left(1-\frac{s}{r^2}\right)^{2+k} \left(1+\frac{2s}{r^2}\right)\,.
\end{eqnarray}
Along the circle one can use the perturbative expansion of $\Pi(s)$ from the
operator product expansion (OPE) to calculate the moments theoretically,
provided $r$ is large enough. As our weight
function $w_r^{k}(s)$ we have chosen a weight function which appears naturally
in $\tau$-decays. This facilitates the inclusion of the isovector $I=1$
contributions from $\tau$-decays which we will use in the second part of our
analysis. It has the important property to suppress the theoretical
contributions at the physical point $s=r^2$. 

The theoretical contributions can be expressed in terms of the Adler function
$D(s)=-s\,d/ds  \Pi(s)$.
In terms of integrals over $D(s)$, the moments ${\mathcal M}^{k}$ take the form 
\begin{center}
\begin{equation}
 \label{eq:2.e}
{\mathcal M}^{k} = - 3\pi i \oint\limits_{|x|=1} \frac{dx}{x} 
\mathcal{F}^{k}(x) D(r^2 x) \,,
\end{equation}
\end{center}
where we have used partial integration and the
$\mathcal{F}^{k}(x)$ can, e.g., be found in Ref. \cite{Gamiz:2002nu}.

Two common approaches exist to evaluate the contour integrals:
in CIPT the exact running
of $\alpha_s(s)$ from the renormalisation group equation along the circle 
is taken into account and the integrals
must be computed numerically. In FOPT the integrand is first expanded in
terms of $\alpha_s(r)$ and the integrals can be calculated analytically. 
In this work we follow the approach of CIPT and comment on FOPT at the end of
this section.


\subsection{Theoretical contributions}

The OPE can be organized in powers of dimension as
\begin{center}
\begin{equation}
\label{eq:2.f}
{\mathcal M}^k(r)=\frac{3}{2}e_s^2 \left(\delta^{Pert}_k+\delta^{m^2}_k
+\delta^{(4)}_k+\delta^{(6)}_k+\delta^{(8)}_k+ ...\right)\,,
\end{equation} 
\end{center}
where the $\delta^{i}_k(r)$ of dimension $i$ are obtained from the contour
integration of the corresponding $D(s)$. The resulting $D$'s can be obtained
from Ref.~\cite{PP2} up to dimension 4.
The perturbative expansion at leading order in the OPE is known to
$O(\alpha_s^3)$. 
The $4^{th}$ order correction was estimated by the method of fastest
apparent convergence (FAC) \cite{Grunberg:1984fw} or the principle of minimal  
sensitivity (PMS) \cite{Stevenson:1981vj} to $\sim (27\pm 50)(\alpha_s/\pi)^4$.
Essential information on ${\mathcal M}(r)$ comes from the dimension 2
contributions as they are leading in the masses. 
Their expansion is known up to $O(\alpha_s^2)$.
For the operators of dimension higher than 4 we have decided to use the values 
given by the ALEPH collaboration in {\cite{Barate98}} rather than the
factorised values since separate discussions of V-A correlators (see, most 
recently {\cite{Cirigliano:2003kc}}) seem to agree with those values rather well.
Furthermore, we have enlarged the error on the dimension 8 condensate by a
factor of 9 to be of the same size as the full contribution which should also
account for the uncertainty from possible higher order contributions.


\subsection{CIPT versus FOPT}

In the context of perturbation theory it seems natural to expand the
theoretical contributions up to a certain power in $\sim\alpha_s^n(r)$ and to
work consistently to that order. This procedure is employed in FOPT where in
the contour integral of eq. \eqn{eq:2.e} the running of $\alpha_s(-s)$ and
$m(-s)$ along the circle is expressed as a power series in $\alpha_s(r)$,
$m(r)$  and 
logarithms of $\ln(-s/r^2)$ up to the desired order. However, as was explained
in \cite{DP}, one gets imaginary logarithms $\ln(-s/r^2)=i(\phi-\pi)$ which are
large in some part of the integration range. So the radius of convergence is
relatively small. It turns out that for an integral containing one power of 
$\alpha_s(-s)$ and for physical values of $\alpha_s(r)$ the
series is just within the convergence radius. However, if the integral contains
higher powers of $\alpha_s^n(-s)$ or $m^n(-s)$ the convergence radius is
smaller and it is not clear whether the expansion can be trusted. The CIPT method
avoids this problematic expansion by solving the renormalisation group equation
of $\alpha_s(-s)$ and  
$m(-s)$ exactly along the circle. At leading order in the OPE the
expansion reads (for $r=1.6\ \gev$ and $k=0$)
\begin{eqnarray}
\label{eq:2.i}
\delta^{Pert}_{CI}&=& 1+0.156+0.032+0.0136+0.0053= 1.207 \nn\\
\delta^{Pert}_{FO}&=&  1+0.168+0.043+0.021+0.0045= 1.24 \,,
\end{eqnarray}
where the $n^{th}$ term includes the contour integral with a power of
$\alpha^n_s(-s)$. 
For FO this corresponds to expanding each integral $A^{(n)}$ in powers up to
$\alpha^4_s(r)$ as in \cite{DP}, eq. (10).
The CIPT shows a better convergence, but the difference
between the two methods is relatively small. When calculating the mass
corrections \cite{PP1}, the situation changes dramatically. We find
\begin{eqnarray}
\label{eq:2.j}
\delta^{m^2}_{CI}/m_s^2(r) &=& -3.366-1.002\nn\\&&
-0.208(+0.66)=-4.58 \nn\\
\delta^{m^2}_{FO}/m_s^2(r) &=&  -3.489-1.398\nn\\&&
-2.158(-2.99)=-7.04 \,.
\end{eqnarray}
In FO the expansion of the corresponding integral $B^{(n)}$ up to powers of
$\alpha^3_s(r)$ can be found in \cite{PP1}, eq. (4.1-4.4).
The final results differ by more than 50\%. The terms in parenthesis of order
$O(\alpha_s^3)$ are estimated from the growth of the coefficient series with FAC
and PMS. We do not include these terms in our final result but use it for the
error estimate. Including these terms both methods would differ by a factor of
2. The improved convergence of the CIPT is obvious. It is interesting to see
how the terms start to diverge with higher powers of $\alpha_s^n(-s)$. Whereas
the difference of the first term is relatively small, the NLO terms already
differ substantially. At order $O(\alpha_s^2)$ they differ by more than a factor
of 10. As explained above, the convergence radius shrinks for higher powers of
the coupling constant and the masses. A discussion on the improved convergence
of CIPT can also be found in \cite{Korner:2000wd}.

Since the $\delta^{m^2}$ are leading in our mass determination they 
directly shift the
result for the strange mass. If we would use FOPT as in 
\cite{Narison95,Maltman98,Maltman:1998xn,Narison99}, our result for $m_s$ would
lower by 25\% up to 40\% with inclusion of the $\alpha_s^3$ estimate.


\section{Phenomenological parametrisation}

We now discuss the phenomenological spectral function. The relevant
contributions for our sum rule are the $\phi(1020)$ and $\phi^{\prime}(1680)$
resonances and a continuum strange quark production. The experimental cross 
section can either be given directly in terms of $R(s)$ or in form of resonance
parameters. 
For the description of the resonances we follow the parametrisation
of Eidelman and Jegerlehner \cite{Eidelman:1995ny}. 

Since the contributions from the $\phi$ constitute an essential input in our
analysis we have included the most recent measurements from CMD-2 \cite{CMD2}.
The product $B_{e^+e^-}B_{K_L^0K_S^0}$ is determined to 
$(1.001\pm0.018)10^{-4}$. Using $B_{K_L^0K_S^0}=0.337\pm0.005$ one obtains
$B_{e^+e^-}=(2.97\pm0.07)10^{-4}$. The corresponding partial width is given in
table \ref{tab:3.a} together with other input parameters of the $\phi$ and
$\phi'$.
\begin{center}
\begin{table}[tbh]
\caption{Resonance parameters for the $\phi$ and $\phi'$.}
\label{tab:3.a}
{\footnotesize
\begin{center}
\begin{tabular}{|c|ccc|}\hline
{} &{} &{} &{} \\[-1.5ex]
{} & $M$ (MeV) & $\Gamma^{ee}$ (keV) & $\Gamma^{tot}$ (MeV) \\[1ex] \hline
{} &{} &{} &{} \\[-1.5ex]
$\phi$ & 1019.5 & $1.271\pm 0.032$ & $4.280\pm 0.041$\\[1ex] 
$\phi'$ & 1680 & 0.53 & 150 \\[1ex]
\hline
\end{tabular}
\end{center}
}
\vspace*{-13pt}
\end{table}
\end{center}
Finally, there is a continuum contribution from open strange quark production.
This contribution can be estimated from ALEPH data \cite{Barate98,Barate:1997hv}
for $K\bar{K}$ production, but it represents only a small contribution compared
to the $\phi$ and $\phi'$ cross section.


\section{Analysis}


\subsection{The $\phi$ sum rule}

In this section we report on preliminary results of \cite{EJS}.
To choose the sum rule window we must fix the values of $k$ and $r$. For small
values of $k$ the perturbative contribution is dominant and the dependence of
the sum rule on the strange quark is small. Higher $k$ improve the sensitivity
on the mass, however, the expansion in $\alpha_s$ converges more slowly. Thus
we choose $k=2,3$ for the central value of our mass and $1\leq k\leq 4$ for the
error estimate. For the energy we use $1.6\ \gev \leq r\leq 2.0\ \gev$. 
In these ranges for
$k$ and $r$ both the perturbative expansion and the phenomenological
parametrisation are under control and the sum rule is reasonably stable. From
an average we obtain $m_s=139\pm 20\ \mev$ where the error is from the sum rule
window only. In addition, one has uncertainties from the different input
parameters, the most important ones from the theoretical expansion of
$\delta^{m^2}$. Furthermore, also the phenomenological parametrisation and the
higher condensates give significant contributions to the error.
Adding all errors quadratically we finally obtain $m_s (2~\gev) = 139 \pm 31 ~ \mev$.
\begin{figure}[tbh]
\hbox to\hsize{\hss
\includegraphics[width=7cm,angle=-90]{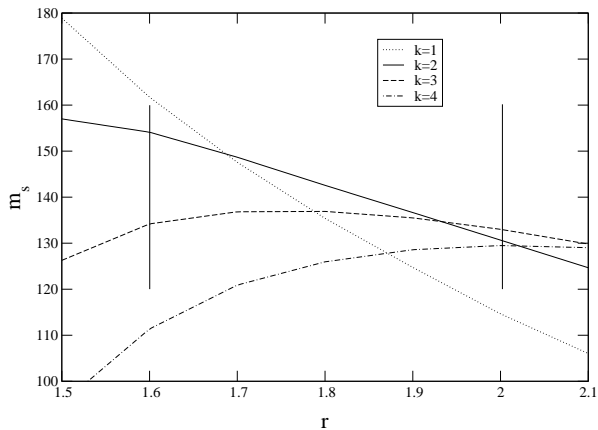}
\hss}
\caption{$m_s$ as function of $r$ for different $k$}
\label{fig:4.a}
\end{figure}


\subsection{The $R_\tau^V-{\mathcal M}^{\phi}$ sum rule}

Here we consider a second sum rule that uses the difference of the
strange vector current and the $\tau\ ud$ vector current which has been
measured in \cite{Barate:1997hv}. The relevant moment reads
\begin{center}
\begin{equation}
 \label{eq:4.b}
{\mathcal M}^{1\phi}=\frac{1}{9} R^V_\tau-{\mathcal M}^\phi
\end{equation}
\end{center}
where ${\mathcal M}^\phi$ is the moment from the strange quark current as discussed
in the last section and $R^V_\tau$ is obtained from the $\tau$ vector
current $j(x)= \bar{u}(x)\gamma_\mu d(x)$.
The advantage of taking this difference is
that the perturbative terms of the OPE drop out, and that the higher
dimensional condensates partially cancel. 
Since the leading order contributions cancel the mass corrections become
leading order. Effectively this sum rules substitutes theoretical uncertainties
by better known experimental values. However, it has the drawback that the data
are only available up to $M_\tau$ and so we can only apply a very limited sum
rule window.
As in the pure strange-vector current sum rule we use values of $k=2,3$ for the
central value and $1\leq k \leq 4$ for the error estimate. Our central mass is
slightly lower with $m_s(2\ \gev)=130\ \mev$ with a similar error as before.
However, since the sum rule window for $r$ is small we cannot check for the
stability of the sum rule over a large range of values. Therefore we take the
strange mass from section 4.1 as our central result and use the 
${\mathcal  M}^{1\phi}$ sum rule as a consistency check.


\section{Conclusions}

The finite energy sum rules provide a powerful method to determine the strange
quark mass from the vector current and our preliminary result is
\begin{center}
\begin{equation}
\label{eq:5.a}
m_s (2~\gev) = 139 \pm 31 ~ \mev\,.
\end{equation}
\end{center}
The theoretical contributions can be
calculated either in FOPT or CIPT. Surprisingly, a large difference is found
and we have commented on this topic in section 2.3. On the phenomenological
side, we have included recent experimental results from CMD-2 \cite{CMD2}.
Our result lies somewhat above other recent determinations 
\cite{Gamiz:2002nu,mutau,Colangelo:1997uy,JOP2,MK}. 
Since this sum rule is independent of other methods, it provides an 
additional and complementary access to the strange quark mass.

\section*{Acknowledgments}
We would like thank Antonio Pich for numerous discussions and reading the
manuscript. Markus Eidem\"uller thanks the European Union for financial 
support under contract no. HPMF-CT-2001-01128. 
This work has been supported in part by 
EURIDICE, EC contract no. HPRN-CT-2002-00311 and by MCYT (Spain) under grant 
FPA2001-3031.

\end{document}